\documentstyle[12pt]{article}
\begin{document}
\vskip 4 cm
\begin{center}
\Large{\bf WHAT THE RIGHT HANDED NEUTRINO REALLY IS?}
\end{center}
\vskip 2 cm
\begin{center}
{\bf Syed Afsar Abbas} \\
Centre for Theoretical Physics\\
JMI, New Delhi - 110025, India\\
(e-mail : afsar.ctp@jmi.ac.in)
\end{center}
\vskip 15 mm  
\begin{centerline}
{\bf Abstract }
\end{centerline}
\vskip 3 mm

We look into the concept of electric charge quantization in the 
Standard Model. The role of the vector nature of electromagnetism and 
that of mass generation by Yukawa coupling is studied. 
We show how the baryon and the lepton numbers arise naturally 
in this picture. This points to an unambiguous and fundamental 
understanding of the actual nature of the right-handed neutrino.
This conforms to Wigner's analysis of the irreducible representations
of the Poincare group.

\newpage

It has been a folklore (clearly reflected even in particle physics 
textbooks) up to the 1990's ( and even sometimes stated today [1] )
that the electric charge is not quantized in the Standard Model (SM)
of particle physics. Hence, it was considered as one of the shortcomings 
of the SM, prompting physicists to go to Grand Unified Theories (GUTs), 
wherein it was shown that electric charge is actually quantized. 
In fact this was considered as the first success of the GUTs concept.
It was only in the late 1990's that it was demonstrated that the electric
charge is actually quantized in the SM itself [2,3,4,5]. 
This demonstration of the electric charge quantization in the SM, 
requires the whole machinery 
of the SM and not just only the group structure, which for SM is 
${SU(3)_{c}} \otimes {SU(2)_{L}} \otimes {U(1)_{Y}}$, but also
the particle content of each generation, spontaneous symmetry breaking 
by a Higgs doublet and anomaly cancellation.

In the present work, we study, as to what it means for the electric 
charge to be right handed and left handed as to the vector nature of 
electromagnetism and as to  how mass is generated by Yukawa coupling 
vis-a-vis electric charge quantization in the SM. 
We show how the concept of the baryon and the lepton numbers arise 
naturally in this picture. This allows us to study the intrinsic and basic 
nature of the right-handed neutrino. This makes connection with the 
irreducible representations of the Poincare group.

Let us first ignore the right-handed neutrino, in say the first generation
of particle, in the SM as represented by the group 
${SU(3)_{c}} \otimes {SU(2)_{L}} \otimes {U(1)_{Y}}$

\begin{displaymath}
q_L = \pmatrix{u \cr d}_L ; (3, 2, Y_q) \end{displaymath} 

\begin{displaymath} u_R ; (3, 1, Y_u)  \end{displaymath}

\begin{displaymath} d_R ; (3, 1, Y_d)  \end{displaymath}

\begin{displaymath} l_L = \pmatrix{\nu  \cr e}_L ; (1, 2, Y_l)  
\end{displaymath}

\begin{equation} e_R ; (1, 1, Y_e)  \end{equation}

Let us now define the electric charge in the most general way in terms of 
the diagonal generators of ${SU(2)_{L}} \otimes {U(1)_{Y}}$ as
\begin{equation}
Q' = a' I_3 + b' Y
\end{equation}
\newline We can always scale the electric charge once as
\begin{equation}
Q = I_3 + b Y
\end{equation}
where $b = {b' \over a'}$

In the SM 
${SU(3)_{c}} \otimes {SU(2)_{L}} \otimes {U(1)_{Y}}$
is spontaneously broken through Higgs mechanism to the group
${SU(3)_{c}} \otimes {U(1)_{em}}$.
Here the Higgs is assumed to be a doublet $\phi$ with arbitrary hypercharge 
$Y_{\phi}$. 
The isospin $I_3 =- {1\over2}$ component of the
Higgs field develops a nonzero vacuum expectation value $<\phi>_o$. Since 
we want
the $U(1)_{em}$ generator Q to be unbroken we require $Q<\phi>_o=0$. This
right away fixes b in (3) and we get

\begin{equation} 
Q = I_3 + ({1 \over 2Y_\phi}) Y 
\end{equation}

To proceed further one imposes the anomaly cancellation conditions
to establish constraints on the various hypercharges above.
First ${[SU(3)_c]}^2 U(1)_Y$ gives $2 Y_q = Y_u + Y_d$
and  ${[SU(2)_L]}^2 U(1)_Y$ gives $3 Y_q = - Y_l$. Next 
${[U(1)_Y]}^3$ does not provide any new constraints. 
So the anomaly 
conditions themselves are not sufficient to provide quantization of 
electric charge in the SM. One has to provide new physical inputs to 
proceed further. There are two independent ways to do so.

{\bf Method 1}:
Here one demands that fermions acquire masses through Yukawa coupling in 
the SM [6]. This brings about the following constraints [4,5]:

\begin{displaymath}
Y_u = Y_q + Y_{\phi}
\end{displaymath}
\begin{displaymath}
Y_d = Y_q - Y_{\phi}
\end{displaymath}
\begin{equation}
Y_e = Y_l - Y_{\phi}
\end{equation}

Note that  $2 Y_q = Y_u + Y_d$ from the anomaly cancellation condition for
${[SU(3)_c]}^2 U(1)_Y$ is automatically satisfied here from the Yukawa
condition above. Now using $3 Y_q = - Y_l$ from anomaly cancellation
along with Yukawa terms above in 
${[U(1)_Y]}^3$ does provide a new constrains of $Y_l = - Y_{\phi}$. 
Putting all these together one immediately gets charge quantization in the 
SM as follows:

\begin{displaymath}
 q_L = \pmatrix{u \cr d}_L , Y_q = {{Y_\phi} \over 3},
\end{displaymath}
\begin{displaymath} Q(u) = {1\over 2} ({1+{1\over 3}}), 
                    Q(d) = {1\over 2} ({-1+{1\over 3}})
\end{displaymath}
\begin{displaymath} u_R, Y_u = {Y_\phi} ({1+{1\over 3}}),
                    Q(u_R) ={1\over 2} ({1+{1\over 3}}) 
\end{displaymath}
\begin{displaymath} d_R, Y_d = {Y_\phi} ({-1+{1\over 3}}),
                    Q(d_R) ={1\over 2} ({-1+{1\over 3}}) 
\end{displaymath}
\begin{displaymath} l_L = \pmatrix{\nu \cr e}_L , Y_l = -Y_\phi,
Q(\nu) = 0, Q(e) = -1
\end{displaymath}
\begin{equation}
 e_R, Y_e = -2Y_\phi, Q(e_R) = -1
\end{equation}

Note that in the above quantization of the electric charge, 
Higgs hypercharge $Y_{\phi}$ always cancels 
out and hence remains unconstrained.
A repetitive structure gives charges for the other generation
of fermions as well.

{\bf Method 2}:
Next we ignore Yukawa coupling and impose the vector nature
of the electric charge [7] which means that photon couples identically to 
the left handed and the right handed charges. That is $Q_L = Q_R$

\begin{displaymath}
{1\over 2} ({1+{{Y_q}\over {Y_\phi}}}) =  
{1\over 2}  {Y_u \over {Y_\phi}}                   
\end{displaymath}
\begin{displaymath}
giving:  Y_u = Y_q + Y_{\phi}
\end{displaymath}
\begin{displaymath}
Q(d) = {1\over 2} ({-1+{{Y_q}\over {Y_\phi}}})=
{1\over 2}  {Y_d \over {Y_\phi}}
\end{displaymath}
\begin{displaymath}
giving: Y_d = Y_q - Y_{\phi}
\end{displaymath}
\begin{displaymath} 
{1\over 2} ({-1+{{Y_l}\over {Y_\phi}}})=
{1\over 2}  {Y_e \over {Y_\phi}}
\end{displaymath}
\begin{equation} 
giving: Y_e = Y_l - Y_{\phi}
\end{equation}

And thereafter charge quantization as in method 1.

Note that the vector condition $Q_l = Q_R$ imposes exactly the same 
constraints on hypercharges as does the Yukawa coupling conditions.
Thus the physical information content of these two methods is exactly the 
same. What does it mean?

We take the vector nature of electromagnetism as obvious. However note 
that 
this is a highly non-trivial property of electromagnetism. For example 
the generator of $U_Y$ does not have it! We take the property of vector 
nature as more fundamental than the Yukawa coupling which are believed to 
be unmotivated and arbitrary in the SM [6]. However from here we 
note that Yukawa coupling is no less fundamental than the vector nature of 
electromagnetism, as the information content vis-a-vis electric charge 
quantization, is exactly the same.

Next, note that for the left handed charge of u- ( similarly for the 
d-quark) one obtained: 
${1\over 2} ({1+{{Y_q}\over {Y_\phi}}}) = 
{1\over 2} ({1+ {1\over 3}})$. Note that $1 \over 3$ is baryon number 
[4,5].
We suggest that this is a general property and that we should define the 
baryon number in SM as  

\begin{equation}
{Y_q \over Y_\phi} = B
\end{equation}

Here in the SM the baryon number is arising as the ratio of the 
hypercharge of the left handed quark with respect to the Higgs 
hypercharge.
Since Higgs is providing the ubiquitous background uniform structure in
which the particles exist, this is a reasonable definition of baryon  
charge for the left handed quarks.

By definition the baryon number is arising for the left handed quark 
representation. What about right handed quarks, In that case we see that
${Y_u \over Y_\phi} = B+1$ for the right handed u-quark and 
${Y_d \over Y_\phi} = B-1$ for the right handed d-quark. 
If one were to define right handed baryon numbers from here then one faces 
the problem that the u- and the d- quarks have different baryon numbers.
Hence the baryon number should arise only for the left handed 
representation. It is important to note that this so called global 
quantum number is chiral in nature. 
This is fine, as the weak interaction has handedness associated with it
and in vector theories both the left and the right handed quarks are 
present in equal proportion. 

Similarly for the left handed electron the electric charge is
${1\over 2} ({-1+{Y_l\over {Y_\phi}}})$
We now associate lepton number with

\begin{equation} 
{Y_l \over {Y_\phi}} = - L 
\end{equation}

which gives the correct charges. Notice that
this is a natural definition of lepton number. Just as for baryon number
in the SM the lepton number arises as the ratio of the hypercharge of the 
left handed lepton with respect to the Higgs hypercharge and as such
is natural to treat this as the lepton number. This more so, as the Higgs
hypercharge remains unconstrained by the theory and the lepton number is 
thus fixed by the background Higgs.
And as for the case of the baryon number, the lepton number is also only 
defined from the left handed representation and is chiral in nature.

Let us now add the right handed neutrino for the first generation 
irreducible representation given in eqn.(1).
Let it be defined as ( in the same notation ): 
\begin{equation}
\nu_R ; (1,1,Y_\nu)
\end{equation}

This brings in additional term from the Yukawa coupling given 
in eqn.(5) as

\begin{equation}
Y_\nu = Y_l + Y_\phi
\end{equation}

Now ${[U(1)_Y]}^3$ anomaly condition, with all the Yukawa couplings and 
the $3 Y_q = - Y_l$ condition, does not provide any new constraint on 
hypercharges. Only that it is consistent with the other conditions.
Without $\nu_R$ it was this anomaly cancellation condition that gave 
crucial information which ensured charge quantization. Now with the 
incorporation of $\nu_R$, as well known [2], the property of electric 
charge quantization is lost.

Thus, let us now impose an empirical constraint. Let us assume that the
the electric charge of this new entity - $\nu_R$ is zero. This would be 
consistent with the overall empirical reality, as any charged $\nu_R$ 
would have 
made its presence felt in laboratory or in cosmological data.
Thus we are demanding it to be inert. We find that
 
\begin{equation}
Q(\nu_R) = {{Y_\nu} \over {2 {Y_\phi}}}
\end{equation}

Note that demanding that
$Q(\nu_R) = 0$ means that it is
${{Y_\nu} \over {Y_\phi}} = 0$. This means that either $Y_\nu = 0$
or ${Y_\phi} = \infty$. Now  ${Y_\phi} = \infty$ is ruled out as
we saw earlier that all the electric charges had factors like
${Y \over {Y_\phi}}$ with Y's for different representations of fermions,
and which will get messed up with this value of $Y_\phi$. 
Hence necessarily:
$Y_\nu = 0$.

What is the significance of this result. We saw earlier that it is the 
ratio of hypercharges with that of the Higgs hypercharge, that the baryon 
number and the lepton number, and the thereon the electric charges get 
defined in the SM. So $Y_\nu = 0$ is special. 
So right away one sees that we cannot define any lepton 
number for this $\nu_R$. So though the left handed neutrino exists and is
identified by its lepton number which puts it in the left handed doublet 
with the electron, the so called right handed neutrino
is completely unlike it, and has no associated lepton number.
Similarly for each generation, however, we may take just a single
$\nu_R$ for all these.

Once we have realized that $\nu_R$ has no lepton number, one sees that
hence there cannot be any Dirac mass (Yukawa coupling) or Majorana mass 
for the left handed neutrino. 
( So if it has mass, some other source has to be 
found or one may be able to explain the oscillation problem through some 
other means ). This $\nu_R$ is colourless, massless, chargeless, 
weak-isospin-less, hypercharge-less and lepton-number-less. 
What is it then?

To understand it we go to Wigner's analysis [8] of the irreducible 
representation of massless entities for the Poincare group.
For massless fields, he showed that also when parity 
is conserved (so for photon), that there are two states of 
polarization +h and -h [9], where h stands for helicity. But for massless 
entities, when parity is not
conserved ( as in the case of weak interaction ), the two states
+h and -h are actually two different irreducible representations
of the Poincare group. The left handed neutrino gets lumped with the
left handed electron by virtue of having a lepton number.
And therefore the other entity, the so called right handed neutrino, 
being of a different representation, is actually quite different.
And this is exactly what we have found here.
Thus our work is a confirmation of Wigner's work on Poicare group.
Clearly so far, the so called right handed neutrino has been 
misunderstood completely.

\newpage
\begin{center}
{\bf REFERENCES }
\end{center}

\vskip 1.2 cm

1. N D Christensen and R Schrock, Phys. Rev. D 72 (2005) 035013

\vskip 1.2 cm

2. R E Marshak, "Conceptual Foundations of Modern Particle Physics",
                World Scientific, Singapore, 1993

\vskip 1.2 cm

3. A Minahan, P Ramond and R C Warner, Phys. Rev. D 41 (1990) 715

\vskip 1.2 cm

4. A Abbas, Phys. Lett B 238 (1990) 344

\vskip 1.2 cm

5. A Abbas, J. Phys. G 16 (1990) L163

\vskip 1.2 cm

6. C Quigg, Rep. Prog. Phys. 70 (2007) 1019

\vskip 1.2 cm

7. S Barr and A Zee, J. Math. Phys. 22 (1981) 2263

\vskip 1.2 cm 

8. E Wigner, Ann. Maths. 40 (1939) 149

\vskip 1.2 cm

9. A Das, "Lectures on Quantum Field Theory", 
          World Scientific, Singapore, 2008

\end{document}